\documentclass[3p,twocolumn]{elsarticle}

\usepackage{verbatim}
\usepackage{geometry}
\usepackage[T1]{fontenc}
\usepackage[utf8]{inputenc}
\usepackage[american]{babel}
\usepackage{epsfig}
\usepackage{graphicx,subcaption,caption}
\usepackage{booktabs}
\usepackage{multirow}
\usepackage{dcolumn}
\usepackage{amsmath}
\usepackage{mathtools}
\usepackage{amsfonts}
\usepackage{amssymb}
\usepackage{epstopdf}
\usepackage{bm}
\usepackage{siunitx}
\usepackage{braket}
\usepackage{enumitem}
\usepackage{soul}
\usepackage[table]{xcolor}
\usepackage{color}
\usepackage{transparent}
\usepackage{pifont}
\usepackage{orcidlink}
\biboptions{sort&compress}
\usepackage[normalem]{ulem}
\usepackage[capitalise]{cleveref}
\crefname{subsection}{Subsection}{Subsections}
\Crefname{subsection}{Subsection}{Subsections}

\definecolor{navyblue}{rgb}{0.0, 0.0, 0.5}
\definecolor{royalblue}{rgb}{0.25, 0.41, 0.88}
\definecolor{cadmiumgreen}{rgb}{0.0, 0.42, 0.24}
\definecolor{blue-violet}{rgb}{0.54, 0.17, 0.89}
\definecolor{darkviolet}{rgb}{0.58, 0.0, 0.83}
\definecolor{orange(colorwheel)}{rgb}{1.0, 0.5, 0.0}

\usepackage{hyperref}
\hypersetup{
    colorlinks=true, 
    linkcolor=darkviolet, 
    citecolor=darkviolet}

\newcommand\mathcomma{\,,}
\newcommand\mathperiod{\,.}

\usepackage{booktabs}
\usepackage{multirow}
\usepackage{dcolumn}
\usepackage{colortbl}

\begin{document}

\begin{frontmatter}

\title{Resilience and implications of adiabatic CMB cooling}

%
\author[1]{Ruchika
\corref{CorrAuthor}}
\ead{ruchika.science@usal.es}
\author[2]{William Giar\`e}
%
\author[3]{Elsa M. Teixeira}
%
\author[4]{Alessandro Melchiorri}
%
%
\address[1]{Departamento de Física Fundamental and IUFFyM, Universidad de Salamanca, E-37008 Salamanca, Spain}
\address[2]{School of Mathematical and Physical Sciences, University of Sheffield, Hounsfield Road, Sheffield S3 7RH, United Kingdom}
\address[3]{Laboratoire Univers \& Particules de Montpellier, CNRS \& Université de Montpellier UMR-5299, 34095 Montpellier, France}
\address[4]{Physics Department and INFN, Universit\`a di Roma ``La Sapienza'', Ple Aldo Moro 2, 00185, Rome, Italy}
\cortext[CorrAuthor]{Corresponding author}
%

\date{\today}

\begin{abstract}
We investigate potential deviations from the standard adiabatic evolution of the cosmic microwave background (CMB) temperature, $T_{\rm CMB}(z)$, using the latest Sunyaev-Zeldovich (SZ) effect measurements and molecular line excitation data, covering a combined redshift range of $0 < z \lesssim 6$. We follow different approaches. First, we reconstruct the redshift evolution of $T_{\rm CMB}(z)$ in a model-independent way using Gaussian Process regression. The tightest constraints come from SZ measurements at $z < 1$, while molecular line data at $z > 3$ yield broader uncertainties. By combining both datasets, we find good consistency with the standard evolution across the full analysed redshift range, inferring a present-day CMB monopole temperature of $T_0 = 2.744 \pm 0.019$ K. Next, we test for deviations from the standard scaling by adopting the parameterisation $T_{\rm CMB}(z) = T_0(1+z)^{1-\beta}$, where $\beta$ quantifies departures from adiabaticity, with $\beta = 0$ corresponding to the standard scenario. In this framework, we use Gaussian Process reconstruction to test the consistency of $\beta = 0$ across the full redshift range and perform $\chi^2$ minimisation techniques to determine the best-fit values of $T_0$ and $\beta$. In both cases, we find good consistency with the standard temperature-redshift relation. The $\chi^2$-minimisation analysis yields best-fit values of $\beta = -0.0106 \pm 0.0124$ and $T_0 = 2.7276 \pm 0.0095$ K, in excellent agreement with both $\beta = 0$ and independent direct measurements of $T_0$ from FIRAS and ARCADE. We discuss the implications of our findings, which offer strong empirical support for the standard cosmological prediction and place tight constraints on a wide range of alternative scenarios of interest in the context of cosmological tensions and fundamental physics.
\end{abstract}

\end{frontmatter}

\section{Introduction}

The cosmic microwave background (CMB) is a cornerstone of modern cosmology, providing a wealth of information about the early Universe, its subsequent evolution, and the initial fluctuations that seeded the large-scale structure of the Universe.

One of the key physical properties of the CMB is the monopole temperature, $T_0$, which represents the current average blackbody temperature of the Universe. The FIRAS instrument~\cite{Fixsen:1996nj} onboard the COBE satellite measured this temperature with high precision to be $T_0 = 2.7255 \pm 0.0006\, \mathrm{K}$~\cite{Fixsen:2009ug}, establishing a critical baseline for the standard $\Lambda$CDM cosmological model.  

A second fundamental physical property of the hot Big Bang theory is that, as the Universe expands, the CMB photon temperature must evolve over cosmic time, cooling down all the way to the present-day CMB monopole temperature $T_0$. Assuming a blackbody spectrum and adiabatic photon cooling in an expanding universe~\cite{Peebles:1994xt}, the CMB temperature evolves with redshift simply as
\begin{equation} 
T_{\rm CMB}(z) = T_0(1+z) \mathperiod 
\label{eq:tz_scal} 
\end{equation}
This behaviour makes the CMB a powerful thermometer of the Universe, as by measuring the present-day monopole temperature, we can infer the temperature of the CMB across the whole cosmic history. However, despite the inherent simplicity and resilience of the temperature-redshift relation in \cref{eq:tz_scal}, deviations from the standard $(1+z)$ scaling could arise in a plethora of different classes of new physics scenarios, mostly pointing to a breakdown of adiabaticity or photon number conservation. While such effects must preserve the excellent agreement with a blackbody spectrum without distorting the CMB beyond the limits set by FIRAS, different temperature-redshift scalings could arise in models with energy injection or removal processes, such as interactions between photons and axion-like particles, decaying vacuum energy, or modified gravity theories (with no claim of completeness, see, \textit{e.g.}, Refs.~\cite{Ellis:2013cu,Bassett:2003vu,Santana:2017zvy,Azevedo:2021npm,Cai:2015emx,Cipriano:2024jng,Avgoustidis:2010ju,Csaki:2001yk,Bassett:2003zw,Chen:1994ch,Cillis:1996qy,Raffelt:1987im,Menard:2009yb,Khoury:2003aq,Khoury:2003rn,Burrage:2007ew,Barrow:1999is,Lee:2021xwh,Goncalves:2019xtc}).

Building on these premises, it is clear that precise measurements of the CMB temperature across redshifts provide a direct yet complementary probe of the Universe, not only tracing its evolution across different epochs but also offering valuable tests of new physics beyond the standard cosmological paradigm. This is also substantiated by the growing interest in exploring potential deviations in $T_{\rm CMB}(z)$. Beyond the fundamental tenet of scientific inquiry that "\textit{Science must begin with myths, and with the criticism of myths} "\footnote{This quotation, from Karl Popper's book "Conjectures and Refutations: The Growth of Scientific Knowledge" \cite{popper2002conjectures}, acknowledges that science is often plagued by "incorrect" understandings of the world (myths) but makes progress through deliberate and systematic efforts to eliminate such misunderstandings (criticism of myths).}, this drive is also practically fuelled by unresolved anomalies and tensions in cosmology, which call for further investigation. 

One such anomaly is the ARCADE (Absolute Radiometer for Cosmology, Astrophysics, and Diffuse Emission) radio excess \cite{Fixsen:2009xn, Seiffert:2011}, which reported an unexpected enhancement in the cosmic radio background at low frequencies. 
While the ARCADE extragalactic spectrum measurements at higher frequencies ($10$, $30$ and $90\, \mathrm{GHz}$) are consistent with the CMB temperature determined by FIRAS (measured at frequencies above $60\, \mathrm{GHz}$), the data shows a clear excess at lower frequencies ($3$ and $8\, \mathrm{GHz}$). This excess is statistically significant, with the $3\, \mathrm{GHz}$ channels being in tension with the FIRAS measurement at more than $5\sigma$ (see Figure 4 of Ref.~\cite{Fixsen:2009xn}). 
The ARCADE analysis finds that the sky-brightness at low frequencies is in good agreement with a CMB monopole temperature of $T_0 = 2.730 \pm 0.004\, \mathrm{K}$ combined with an additional power-law component, characterised by a spectral index of $-2.6$, to account for the observed excess.  
The ARCADE anomaly challenges the standard assumptions of cosmological thermal history, suggesting either the presence of new physics \cite{Fornengo:2011cn} or contributions from previously unaccounted astrophysical sources \cite{Singal:2017jlh}. Precise measurements of $T_{\rm CMB}(z)$ at low redshifts are therefore essential to determine whether the ARCADE excess could be linked to deviations from the standard cooling of the CMB \cite{deMartino:2015ema} or if it arises from non-cosmological sources such as galactic or extragalactic emissions \cite{Subrahmanyan:2013eqa}. 

Another key challenge is the Hubble tension  ~\cite{Verde:2019ivm,DiValentino:2020zio,DiValentino:2021izs,Perivolaropoulos:2021jda,Schoneberg:2021qvd,Shah:2021onj,Abdalla:2022yfr,DiValentino:2022fjm,Kamionkowski:2022pkx,Giare:2023xoc,Hu:2023jqc,Verde:2023lmm,DiValentino:2024yew,Perivolaropoulos:2024yxv,DiValentino:2025sru}, the persistent discrepancy between the locally measured Hubble constant, $H_0$, from distance-ladder-based methods, most notably the one derived by the SH$_0$ES collaboration~\cite{Riess:2021jrx}, and the value inferred from the CMB observations by the \textit{Planck} satellite under $\Lambda$CDM~\cite{Planck:2018vyg}. The local measurements consistently yield a higher value of $H_0$ compared to the \textit{Planck} predictions, with the tension now exceeding $5\sigma$. This discrepancy could point to unknown physical processes, such as modifications to the expansion history of the Universe, interactions between dark energy and matter, or early-time new physics\footnote{With no claim of completeness, for discussions in these directions see, \textit{e.g.}, Refs.~\cite{Anchordoqui:2015lqa,Karwal:2016vyq,Benetti:2017juy,Mortsell:2018mfj,Kumar:2018yhh,Guo:2018ans,Poulin:2018cxd,Graef:2018fzu,Agrawal:2019lmo,Escudero:2019gvw,Niedermann:2019olb,Sakstein:2019fmf,Knox:2019rjx,Hart:2019dxi,Ballesteros:2020sik,Jedamzik:2020krr,Ballardini:2020iws,DiValentino:2020evt,Niedermann:2020dwg,Gonzalez:2020fdy,Braglia:2020auw,RoyChoudhury:2020dmd,Brinckmann:2020bcn,Karwal:2021vpk,Herold:2022iib,Gomez-Valent:2021cbe,Cyr-Racine:2021oal,Niedermann:2021ijp,Saridakis:2021xqy,Herold:2021ksg,Odintsov:2022eqm,Aboubrahim:2022gjb,Ren:2022aeo,Adhikari:2022moo,Nojiri:2022ski,Schoneberg:2022grr,Joseph:2022jsf,Gomez-Valent:2022bku,Odintsov:2022umu,Ge:2022qws,Schiavone:2022wvq,Brinckmann:2022ajr,Khodadi:2023ezj,Kumar:2023bqj,Ben-Dayan:2023rgt,Ruchika:2023ugh,Yadav:2023yyb,Sharma:2023kzr,Ramadan:2023ivw,Fu:2023tfo,Efstathiou:2023fbn,Montani:2023ywn,Stahl:2024stz,Vagnozzi:2023nrq,Zhai:2023yny,Garny:2024ums,Co:2024oek,Toda:2024ncp,Giare:2024ytc,Teixeira:2025czm} or Refs.~\cite{Abdalla:2022yfr,DiValentino:2025sru} for recent reviews.}. Since the evolution of $T_{\rm CMB}(z)$ directly probes the thermal history of the Universe, testing deviations from the standard scaling and constraining the present-day CMB monopole temperature in a model-independent way could shed light on the origin of the Hubble tension itself. To illustrate this point better, we can consider two representative scenarios. Let us first consider the case in which the standard scaling given by \cref{eq:tz_scal} holds. In this case, $T_0$ remains -- at its core -- another free parameter of $\Lambda$CDM (as well as all its extensions). While $T_0$ is typically fixed to the FIRAS best-fit value in standard CMB analyses, relaxing this prior and allowing $T_0$ to vary freely introduces a strong degeneracy with other cosmological parameters, most notably $H_0$.\footnote{The reason is that, from a CMB perspective, the net effect of $T_0$ is to quantify the amount of expansion between a given cosmic epoch and today. Different values of $T_0$ translate into different angular diameter distances to the last scattering surface; an effect that is almost perfectly degenerate with changes in $H_0$.} Therefore, as shown in~\cite{Ivanov:2020mfr}, in principle, the Hubble tension can be reinterpreted as a $3\sigma$ discrepancy between the value of $T_0$ that can be inferred from \textit{Planck} CMB data combined with a SH$_0$ES prior on $H_0$ and the value directly measured by FIRAS.\footnote{It's worth noting that combining \textit{Planck} with Baryon Acoustic Oscillation (BAO) measurements restores consistency with FIRAS, see, \textit{e.g.}, Refs.~\cite{Ivanov:2020mfr,ACT:2025tim}.} Moving instead beyond the standard scaling given by \cref{eq:tz_scal} opens the door to a much richer phenomenology, allowing for physical mechanisms that directly modify the adiabatic cooling of the Universe, possibly affecting the inferred value of $H_0$. For instance, as discussed later in this work, modifications to the temperature–redshift relation can lead to violations of the cosmic distance duality relation (DDR) \cite{Etherington1933,Ellis2007}, which connects the angular diameter distance and the luminosity distance. As recently argued in Ref.~\cite{Teixeira:2025czm}, such violations can alter the calibration of cosmic distances and offer a novel way to explain the Hubble tension.
 
From an observational standpoint, $T_{\rm CMB}(z)$ has been measured using two key methods: the Sunyaev-Zeldovich (SZ) effect in galaxy clusters \cite{Luzzi:2009ae, Hurier:2013ona} and molecular line excitations in distant astrophysical environments \cite{Muller:2012kv, Klimenko:2021wti}. These measurements span a range of redshifts and offer direct constraints on the thermal history of the Universe. While the SZ effect probes the inverse Compton scattering of CMB photons by high-energy electrons in galaxy clusters \cite{Sunyaev:1972eq}, molecular line observations provide independent estimates of $T_{\rm CMB}(z)$ by analysing the excitation states of molecules in distant quasars and other environments \cite{Srianand:2000wu, Noterdaeme:2010tm}. These datasets enable direct reconstructions of $T_{\rm CMB}(z)$, potentially grounded in machine learning techniques, which are useful for model-independent tests of the temperature-redshift scaling. Additionally, they help constrain specific theoretical scenarios that predict deviations from the standard scaling, making them a valuable tool for assessing the validity of various cosmological models \cite{Luzzi:2009ae, Avgoustidis:2011aa}.

In this study, we present the most up-to-date constraints on the evolution of the CMB temperature with redshift. First, we reconstruct the redshift evolution of $T_{\rm CMB}(z)$ in a model-independent way using Gaussian Process (GP) regression techniques. Then, we parameterise deviations from the standard temperature-redshift scaling as~\cite{Lima:2000ay, LoSecco:2001zz, Hofmann:2023wyk, Euclid:2020ojp, SPT:2013gam, Bengaly:2020vly,Avgoustidis:2015xhk}
\begin{equation}
    T_{\rm CMB}(z) = T_0(1+z)^{1-\beta} \mathcomma
    \label{eq:tzbeta_scal}
\end{equation}
where $\beta$ quantifies departures from the standard scaling in \cref{eq:tz_scal}, recovered for $\beta=0$. While this parameterisation can capture a wide range of mechanisms that may lead to non-adiabatic cooling (or heating) processes affecting CMB photons (discussed throughout the manuscript), our primary goal is to take an agnostic approach and test possible deviations from $\beta=0$ over a broad redshift range, using a dataset of $T_{\rm CMB}(z)$ measurements obtained from the SZ effect and molecular line observations \cite{DeBernardis:2006ii}. To do so, we mainly employ GP reconstruction techniques \cite{Seikel:2012uu, Yennapureddy:2017vvb} and derive a smooth representation of $\beta$ that captures both the trends and uncertainties in the data. Focusing on low-redshift $T_{\rm CMB}(z)$ measurements, where observational uncertainties are smaller, we enhance the precision of our results and provide tighter constraints on the temperature-redshift scaling~\cite{Chluba:2014wda, deMartino:2015ema}. Finally, by minimising the $\chi^2$ statistics, we determine the best-fit value of $T_0$ and $\beta$ and its confidence intervals, allowing us to rigorously test the consistency of the $\beta$-parametrised model with the observed data~\cite{Euclid:2020ojp, Avgoustidis:2011aa}.

Our analysis provides a comprehensive and timely update on potential deviations in the CMB temperature evolution, deriving constraints that can be used to test and compare with anomalies such as the ARCADE excess and the Hubble tension while offering valuable insights into the thermal and dynamical history of the universe \cite{Martins:2024okq, Jetzer:2010xe} and supporting model-building efforts in the broader cosmology community.

This paper is organised as follows. \cref{sec:methods_data} outlines the methodology used in this study. In particular, \cref{sec:gauss_proc} introduces the Gaussian Process reconstruction technique used to model the redshift evolution of the CMB temperature, while \cref{sec:data} provides details on the observational data used. \cref{sec:results} presents the results of our analysis. \cref{sec:gp_rec} discusses the model-independent GP reconstruction of the temperature-redshift relation. Following this, \cref{sec:tz_dev} looks at constraints on potential deviations from the standard adiabatic temperature scaling under the modified temperature-redshift relation. For determining the best-fit parameters of temperature evolution, \cref{sec:chi2} employs a $\chi^2$ minimisation approach. Implications of our findings are explored in \cref{sec:discussion}, with specific attention to CMB monopole temperature constraints in \cref{sec:t0_const} and physical mechanisms that could modify the temperature-redshift scaling in \cref{sec:beta_const}. Finally, \cref{sec:conc} concludes by summarising the main results and suggesting future research directions.

\section{Methods and Data} 

\label{sec:methods_data}

\subsection{Gaussian Process Reconstructions} \label{sec:gauss_proc}

We employ GP as a regression tool that allows to infer back some function $f$ given the set of data~\cite{rasmussen2006gaussian}. Any given GP reconstructed function $f$ depends mainly on the choice of the kernel (or covariance matrix) $k$ and mean prior functions $\mu$~\cite{Holsclaw:2010sk, Holsclaw:2011wi}. These functions define the properties of the underlying process being modelled and incorporate prior knowledge and physical assumptions about the underlying physical process. Therefore, they play a critical role in shaping the behaviour and predictions of the GP reconstructions~\cite{Williams1998,Seikel:2013fda, Yennapureddy:2017vvb}. 

In more quantitative terms, $f$ can be expressed in terms of $k$ and $\mu$ as
\begin{equation}
f \sim {N(\mu(z),k(z_i,z_j))} \mathcomma
\end{equation}
where $k(z_i,z_j)$ is the covariance matrix element $(i,j)$ at the input points $\{z_i,z_j\}$ and the notation $N(\cdot)$ indicates that the GP is evaluated at points drawn from a normal distribution~\cite{rasmussen2006gaussian}. Therefore, two important choices have to be made when it comes to GP reconstructions. 
\begin{itemize}
    \item The covariance matrix (or kernel) $k$: under GP reconstruction, the covariance term is the term that defines similarity or nearness to the actual function. Choosing some specific covariance matrix or kernel can give different results once the covariance matrix is changed \cite{Liao:2019qoc}, possibly introducing bias in the results, see \textit{e.g.}, Refs.~\cite{Keeley:2019hmw, Mukherjee:2018ebj, Mukherjee:2020hyn}. We employ the squared exponential (SE) covariance function, which has the form: 
    \begin{equation} 
    k_{\mathrm{SE}} (r) = \sigma_f^2 \exp \left( \frac{-r^2}{l_f^2} \right) \mathcomma
    \end{equation} 
    where $\sigma_f$ is the overall amplitude of the correlation, $l_f$ defines the characteristic scale of the coherence length of correlation, and $r=|z_i - z_j |$ is the coherence length (or distance between input points)~ \cite{Seikel:2012uu, Shafieloo:2012ht}.\footnote{This is the most commonly used covariance function in the literature. However, we explicitly checked that our results are not sensitive to the choice of kernel.}

    \item The mean function $\mu(z)$: this is the initial guess of the function. Being one of the key ingredients in reconstructing the GP function, having a good initial guess can help reduce the dynamic range over which the data needs to be fit \cite{Williams1998}. Therefore, caution must be taken in adopting a prior mean function, and it should not be arbitrary \cite{Holsclaw:2011wi}. The general choice of the prior mean function is the zero mean function, as it is the most model-independent choice~\cite{Seikel:2013fda}. We adopt this approach in our work as well\footnote{Note that adopting the prior mean function to be zero may result in needing large values of $\sigma_f$ to bring the function $f$ close to observational datasets $y(z)$~\cite{Yahya:2013xma} as well as large values of $\sigma_f$ to bring $f(z)$ close to $y(z)$ over a large redshift range would also require a large coherence length $r$. For these reasons, many argued that another suitable choice of mean function for cosmological applications could be selecting the standard predictions for the quantity being reconstructed. However, choosing a model-dependent prior function may not necessarily be the best choice because the final results do not turn out to be completely independent of this initial guess. Instead, they retain some memory, potentially biasing the results~\cite{Gomez-Valent:2018hwc}.}.
\end{itemize}

\subsection{Observational Data Used} \label{sec:data}

\begin{table*}[htb]
\footnotesize
\centering 
\renewcommand{\arraystretch}{1.2}
\resizebox{0.8\textwidth}{!}{\begin{tabular}{lccr||lccr}
\toprule
$z$ & $T_{\rm CMB}(z) \pm \sigma_T$ [K] & Type & Ref. & $z$ & $T_{\rm CMB}(z) \pm \sigma_T$ [K] & Type & Ref. \\
\hline \hline
0.037 & $2.888 \pm 0.041$ & \text{SZ} & \cite{Hurier:2013ona}   & 0.718 & $4.933 \pm 0.371$ & \text{SZ} & \cite{Hurier:2013ona} \\
0.072 & $2.931 \pm 0.020$ & \text{SZ} & \cite{Hurier:2013ona}   & 0.742 & $5.01^{+0.49}_{-0.33}$ & \text{SZ} & \cite{SPT:2013gam} \\
0.125 & $3.059 \pm 0.034$ & \text{SZ} & \cite{Hurier:2013ona}   & 0.783 & $4.515 \pm 0.621$ & \text{SZ} & \cite{Hurier:2013ona} \\
0.129 & $3.01^{+0.14}_{-0.11}$ & \text{SZ} & \cite{SPT:2013gam} & 0.870 & $5.356 \pm 0.617$ & \text{SZ} & \cite{Hurier:2013ona} \\
0.171 & $3.197 \pm 0.032$ & \text{SZ} & \cite{Hurier:2013ona}   & 0.887 & $4.97^{+0.24}_{-0.19}$ & \text{SZ} & \cite{SPT:2013gam} \\
0.220 & $3.288 \pm 0.035$ & \text{SZ} & \cite{Hurier:2013ona}   & 0.890 & $5.08 \pm 0.10$ & \text{Spec} & \cite{Muller:2012kv} \\
0.265 & $3.44^{+0.16}_{-0.13}$ & \text{SZ} & \cite{SPT:2013gam} & 0.972 & $5.813 \pm 1.025$ & \text{SZ} & \cite{Hurier:2013ona} \\
0.273 & $3.416 \pm 0.040$ & \text{SZ} & \cite{Hurier:2013ona}   & 1.022 & $5.37^{+0.22}_{-0.18}$ & \text{SZ} & \cite{SPT:2013gam} \\
0.332 & $3.562 \pm 0.052$ & \text{SZ} & \cite{Hurier:2013ona}   & 1.73 & $7.9^{+1.7}_{-1.4}$ & \text{Spec} & \cite{Klimenko:2021wti}\\
0.371 & $3.53^{+0.18}_{-0.14}$ & \text{SZ} & \cite{SPT:2013gam} & 1.77 & $6.6^{+1.2}_{-1.1}$ & \text{Spec} & \cite{Klimenko:2021wti}\\
0.377 & $3.717 \pm 0.065$ & \text{SZ} & \cite{Hurier:2013ona}   & 1.78 & $7.2 \pm 0.8$ & \text{Spec} & \cite{Cui:2005qw}\\
0.416 & $3.82^{+0.19}_{-0.15}$ & \text{SZ} & \cite{SPT:2013gam} & 1.97 & $7.9 \pm 1.0$ & \text{Spec} & \cite{2001ApJ...547L...1G}\\
0.428 & $3.971 \pm 0.073$ & \text{SZ} & \cite{Hurier:2013ona}   & 2.04 & $8.6^{+1.9}_{-1.4}$ & \text{Spec} & \cite{Klimenko:2021wti}\\
0.447 & $4.09^{+0.25}_{-0.19}$ & \text{SZ} & \cite{SPT:2013gam} & 2.34 & $10.0 \pm 4.0$ & \text{Spec} & \cite{Srianand:2000wu}\\
0.471 & $3.943 \pm 0.113$ & \text{SZ} & \cite{Hurier:2013ona}   & 2.42 & $9.04^{+0.9}_{-0.7}$ & \text{Spec} & \cite{Klimenko:2021wti}\\
0.499 & $4.16^{+0.27}_{-0.20}$ & \text{SZ} & \cite{SPT:2013gam} & 2.53 & $9.8^{+0.7}_{-0.6}$ & \text{Spec} & \cite{Klimenko:2021wti}\\
0.525 & $4.380 \pm 0.120$ & \text{SZ} & \cite{Hurier:2013ona}   & 2.63 & $10.8^{+1.4}_{-3.3}$ & \text{Spec} & \cite{Klimenko:2021wti}\\
0.565 & $4.075 \pm 0.157$ & \text{SZ} & \cite{Hurier:2013ona}   & 2.69 & $10.4^{+0.8}_{-0.7}$ & \text{Spec} & \cite{Klimenko:2021wti}\\
0.590 & $4.62^{+0.36}_{-0.26}$ & \text{SZ} & \cite{SPT:2013gam} & 3.02 & $12.1^{+1.7}_{-3.2}$ & \text{Spec} & \cite{Molaro:2001jv}\\
0.619 & $4.404 \pm 0.195$ & \text{SZ} & \cite{Hurier:2013ona}   & 3.09 & $12.9^{+3.3}_{-4.5}$ & \text{Spec} & \cite{Klimenko:2021wti}\\
0.628 & $4.45^{+0.31}_{-0.23}$ & \text{SZ} & \cite{SPT:2013gam} & 3.29 & $15.2^{+1.0}_{-4.2}$ & \text{Spec} & \cite{Klimenko:2021wti}\\
0.676 & $4.779 \pm 0.279$ & \text{SZ} & \cite{Hurier:2013ona}   & 6.34 & $23.1^{+7.1}_{-6.7}$ & \text{Spec} & \cite{Riechers:2022tcj}\\
0.681 & $4.72^{+0.39}_{-0.27}$ & \text{SZ} & \cite{SPT:2013gam} & & & \\
\bottomrule
\end{tabular}}
\caption{Collection of Sunyaev-Zeldovich (SZ) effect and spectroscopic (Spec) measurements of the CMB temperature at various redshifts.}
\label{tab:combined_data}
\end{table*}

\begin{figure}[tb]
    \centering
\includegraphics[width=0.95\columnwidth]{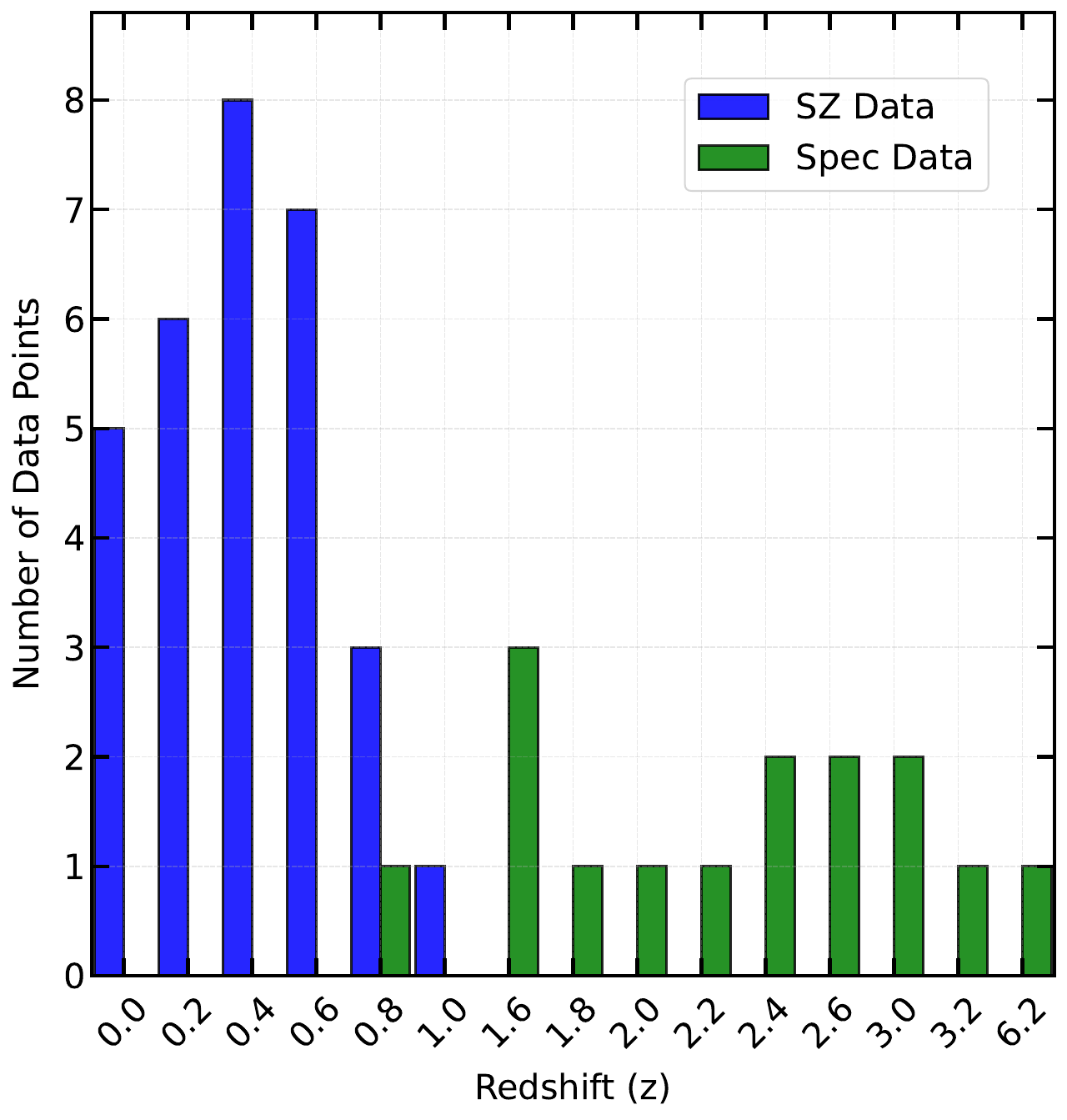}
    \caption{Illustration of the number of Sunyaev-Zeldovich (SZ) effect and spectroscopic (Spec) measurement data points at different redshift bins. For creating a clear visual separation, the placement of bars follows a specific arrangement: the SZ data are placed at the bin centre minus 0.5 bar width and the Spec data at the bin centre plus 0.5 bar width. This strategic offset between the two datasets within each bin allows for easy comparison of data availability across the entire redshift range.}
    \label{fig:cc_data}
\end{figure}

In this study, we use measurements of the CMB temperature obtained \textit{via} two distinct observational techniques:
\begin{itemize} 
\item Thermal Sunyaev–Zeldovich effect (\textbf{SZ}) in galaxy clusters (typically at intermediate redshifts $z\lesssim 1.5$): as the CMB photons travel through hot ionised gas in galaxy clusters their energy will change as they scatter off the free, thermal electrons in the gas through inverse Compton scattering \cite{Sunyaev:1970er,Sunyaev:1972eq}. CMB observatories like \textit{Planck} \cite{Planck:2018nkj}, ACT \cite{ACT:2020gnv} and SPT \cite{SPT-3G:2022hvq} observe the small spectral distortions in the intensity of the CMB around clusters at different frequencies, which can be combined with models of the cluster's gas properties to infer the CMB temperature at the redshift of the cluster \cite{Battistelli:2002ie,Luzzi:2009ae}. 
More precisely, we use a compilation of 30 highly precise measurements of $T_{\rm CMB}(z)$ obtained from the SZ effect \cite{Gelo:2022vsa}, covering the redshift range $0.037 \le z \le 1.022$.
These measurements are derived from a set of 815 \textit{Planck} clusters organised into 18 redshift bins \cite{Hurier:2013ona}, and a sample of 158 SPT clusters organised into 12 redshift bins \cite{SPT:2013gam}.

\item High-resolution spectroscopy (\textbf{Spec}) of molecular species whose energy level populations are sensitive to CMB-induced excitation: some molecules in distant galaxies or quasars have energy levels that can be excited by CMB photons. Radio and sub-millimetre telescopes like ALMA~\cite{2009IEEEP..97.1463W} and NOEMA~\cite{NOEMA} can observe the absorption or emission lines of these molecules (such as CO, CN and H$_2$O). The relative population of different molecular energy levels depends strongly on the surrounding CMB temperature. Analysing the ratios of line intensities makes it possible to estimate the local value of $T_{\rm CMB} (z)$ at the absorber's redshift with precision. However, this method is more effective at higher redshifts (typically $1 \lesssim z \lesssim 3$) where the CMB temperature is higher, and the excitation effects are more pronounced.
In particular, in this work, we analyse a set of 15 measurements based on molecular line excitations obtained through observations across various wavelengths, from optical to radio, using a range of molecular or atomic species (hereafter referred to as spectroscopic data), which extend over a broader and higher redshift interval ($0.89 \le z \le 6.34$), albeit with larger uncertainties \cite{Gelo:2022vsa}. A recent compilation~\cite{Klimenko:2021wti} updates several earlier analyses~\cite{Srianand:2008fz,Noterdaeme:2010gv,Noterdaeme:2010tm,Noterdaeme:2016ykw,Muller:2012kv,Cui:2005qw,2001ApJ...547L...1G,Molaro:2001jv} by accounting for the contribution of collisional excitation in the diffuse interstellar medium to the excitation temperature of the tracer species. 
While the resolution of this technique initially only allowed for upper limits on $T_{\rm CMB} (z)$, continued efforts over the past couple of decades~\cite{Srianand:2000wu} have yielded several measurements, with the compilation used in this work extending up to $z \sim 6.34$~\cite{Riechers:2022tcj} but with most measurements being in the range $z \lesssim 3.5$. However, it does not include measurements with upper limits only due to negligible statistical weight in the sample.
\end{itemize} 

The full dataset is summarised in \cref{tab:combined_data}, with the redshift distribution being illustrated in the histogram in \cref{fig:cc_data}, which offers a clearer view of the redshift range and respective weight. To the best of our knowledge, this compilation of data was first used in Ref.~\cite{Gelo:2022vsa}, where the authors performed a comparative analysis of the cosmological constraining power of CMB temperature measurements, showing that they are broadly competitive with other low-redshift datasets across several different classes of models. In Refs.~\cite{Martins:2024okq,Martins:2025kzd}, the authors extended this study by analysing additional models and forecasting the potential improvements achievable with future high-precision CMB temperature measurements. Conversely, as detailed in the following sections, in this work, we explore the constraining power of this dataset by employing model-independent techniques.

\section{Results} \label{sec:results}

\subsection{Model-independent GP Reconstruction} \label{sec:gp_rec}

\begin{figure*}[htb]
    \centering
    \includegraphics[width=\textwidth]{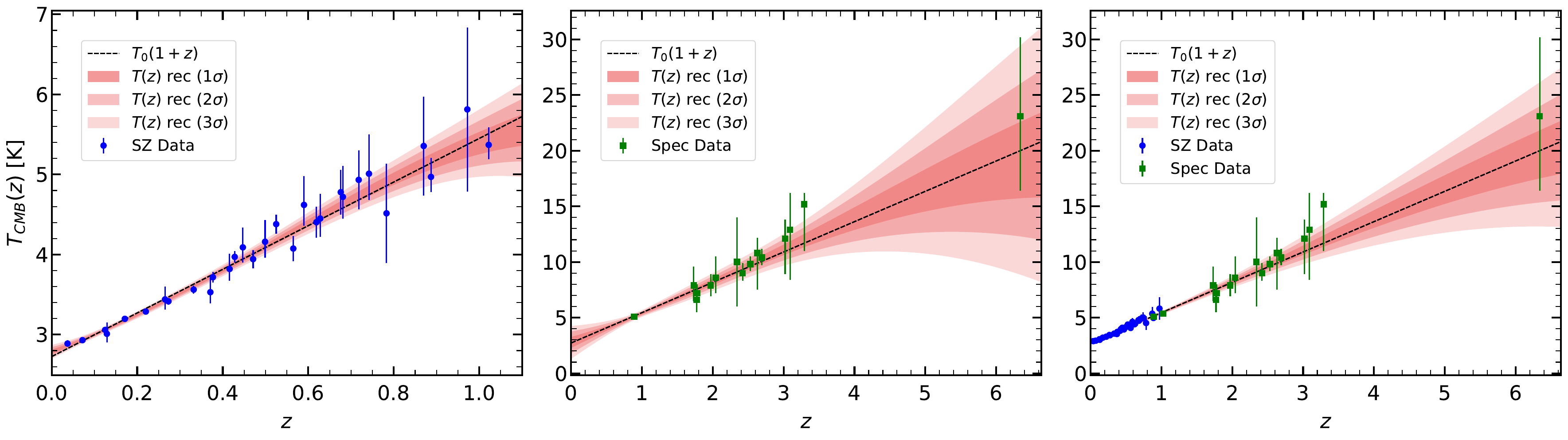}
    \caption{Reconstruction of $T_{\rm CMB}(z)$ through GP as a function of redshift. The increasingly lighter shaded pink regions represent the $1\sigma$, $2\sigma$ and $3\sigma$ confidence intervals. The reconstruction is performed considering only the SZ data (left panel, blue circles) and the spectroscopic measurements (middle panel, green squares), as well as their combination (right panel).}
    \label{fig:cc_data32}
\end{figure*}

The first step of our analysis is to perform a model-independent reconstruction of $T_{\rm CMB}(z)$ as a function of redshift using GP and adopting the squared-exponential kernel. Note that this kernel requires symmetric uncertainties, whereas some of the data points reported in \cref{tab:combined_data} feature asymmetric error bars. We handle these cases by modelling each data point as a shifted Gaussian distribution, where the central value is displaced from the original by half the difference between the positive and negative uncertainties, and the effective symmetric standard deviation is defined as the root-mean-square of the two errors. \footnote{This procedure is essentially a way to "Gaussianise" the original 1D posterior distribution, preserving as much as possible the high-likelihood region without significantly over- or underestimating the tails of the distribution. In practical terms, this method improves upon simply averaging the asymmetric uncertainties. When the error bars are strongly asymmetric, taking their average gives disproportionate weight to the side with the smaller uncertainty and under-represents the side with the longer tail. By contrast, shifting the mean and computing an effective $\sigma$ leads to a Gaussian approximation that better reflects the actual shape of the original distribution, as the mean is naturally pulled toward the wider side, resulting in a more balanced representation of the data's uncertainty.
Nevertheless, we have tested that our results do not significantly depend on this choice. Alternative approaches to treating asymmetric uncertainties yield results that are largely compatible.} 

\cref{fig:cc_data32} illustrates this reconstruction. In the left panel, we show in blue the data points with their respective error bars from SZ measurements, spanning redshifts up to $z \sim 1$. In the middle panel, we show the spectroscopic measurements in green, which extend to higher redshifts $z \sim 6$ but feature larger uncertainties than the SZ data. In the right panel, we show the two datasets combined, maintaining the same colour scheme: blue for the SZ data points and green for the spectroscopic data. In all three panels, the reconstructed function (based on SZ data alone, spectroscopic data alone, and the combination of SZ and spectroscopic data, respectively) is shown together with its $1\sigma$, $2\sigma$, and $3\sigma$ uncertainty bands, indicated by progressively lighter shades of pink. The dashed black line represents the standard adiabatic cooling relation, $T_{\rm CMB}(z) = T_0(1+z)$, where $T_0$ is fixed to the value measured by FIRAS~\cite{Fixsen:2009ug}.

Focusing on the low-redshift SZ-based reconstruction in the left panel, we find good consistency with the standard scaling: the reconstructed curve remains within two standard deviations of the expected $T_{\rm CMB}(z)$ throughout. From our model-independent SZ-based reconstruction of $T_{\rm CMB}(z)$ at $z \lesssim 1$, we can infer the value of $T_0 \equiv T_{\rm CMB}(z=0)$, providing a model-independent estimate of the present-day CMB monopole temperature. We obtain $T_0 = 2.785\pm0.029$ K, which is in good agreement with the value directly measured by FIRAS (ARCADE), with a difference at the level of about $2.0\sigma$ ($1.9\sigma$).\footnote{Here and in what follows, we quantify the consistency between our estimate of the present-day CMB monopole temperature and the direct measurements provided by FIRAS and ARCADE by adopting the so-called "rule of thumb difference in means" which involves comparing the difference in the mean values of $T_0$ obtained from two datasets to the quadrature sum of their uncertainties, as defined in Eq. (40) of Ref.~\cite{Raveri:2018wln}.}

In contrast, the reconstruction based on high-redshift spectroscopic data (middle panel) exhibits broader uncertainty bands, reflecting not only the larger error bars associated with the individual measurements but also the uneven redshift coverage of the dataset. In particular, the reconstruction is well constrained only in the redshift range densely populated by data points, roughly between $z \sim 1$ and $z \sim 3.5$. Beyond this interval, the lack of measurements (especially the absence of data between $z \sim 3.5$ and $z \sim 6$) leads to a noticeable growth in the uncertainty bands. Similarly, the lack of spectroscopic points at $z < 1$ prevents the reconstruction from being precise at low redshifts. As a result, when relying on this dataset alone, we cannot extract a competitive estimate of the present-day CMB temperature $T_0$. That said, the reconstruction remains fully consistent (within one standard deviation) with the standard temperature-redshift scaling \cref{eq:tz_scal} (black dashed line in the figure), and we find no indication of deviations from this expected behaviour.

Finally, when the SZ and spectroscopic data are analysed in combination (right panel), the overall reconstruction remains consistent with the standard temperature-redshift relation within $1\sigma$ across the full redshift range probed by the data. In this case, the complementarity between the densely populated SZ measurements at low redshift and the spectroscopic points at higher redshift not only enhances the constraining power in their respective domains but also improves the overall reconstruction in the redshift intervals that are more sparsely sampled. This is particularly evident when comparing the right panel with the middle one in the same figure, where the uncertainty bands are visibly reduced even in the regions with fewer direct data points. In this case, using the full GP reconstruction of $T_{\rm CMB}(z)$, we obtain
\begin{equation}
T_0 = 2.744 \pm 0.019\, \text{K}\mathcomma
\end{equation}
in very good agreement with the direct measurements from FIRAS (ARCADE) at $\sim 1\sigma$ ($\sim 0.7\sigma$).

\subsection{Beyond the standard adiabatic cooling} 
\label{sec:tz_dev}

\begin{figure*}[htb]
    \centering
\includegraphics[width=\textwidth]{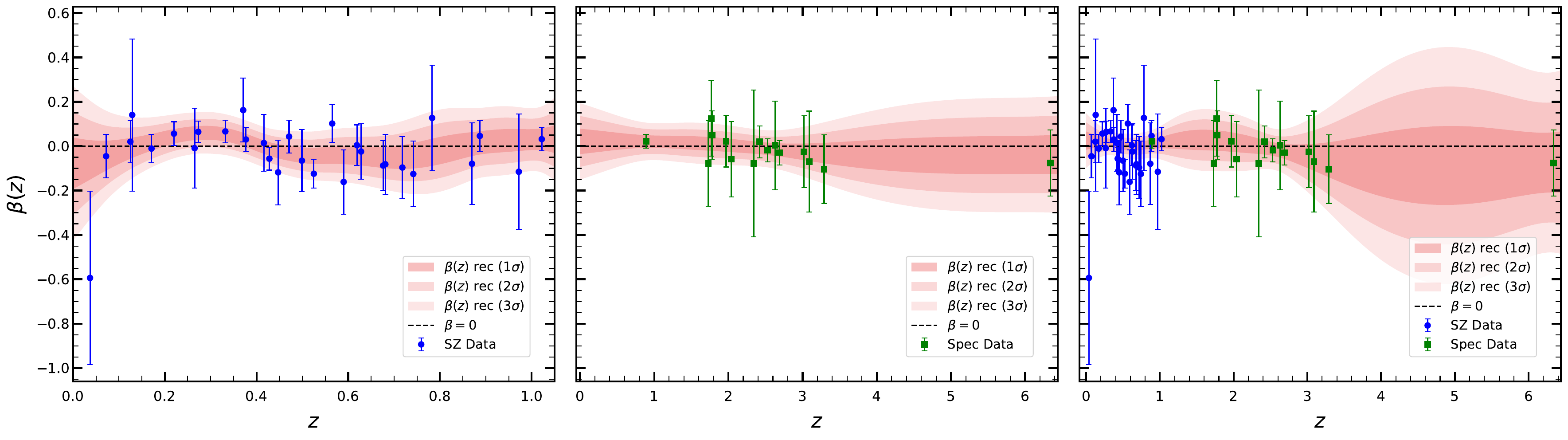}
    \caption{Reconstructed evolution of the deviation parameter $\beta$ as a function of redshift using Gaussian Processes regression applied to the observational CMB temperature datasets. The increasingly lighter pink regions represent the $1\sigma$, $2\sigma$ and $3\sigma$ confidence intervals of the reconstruction, illustrating potential redshift-dependent behaviour of the deviation from the standard cosmological scaling ($\beta=0$). The reconstruction is performed considering only the SZ data (left panel, blue circles) and the spectroscopic measurements (middle panel, green squares), as well as their combination (right panel).}
    \label{fig:cc_data_recbeta}
\end{figure*}

The second step of our analysis is to examine further potential deviations from the standard adiabatic temperature scaling, which can be quantified by introducing 
the modified temperature-redshift relation given by \cref{eq:tzbeta_scal}. This equation can be inverted as
\begin{equation}
\beta (z) = 1 - \frac{\ln \left( T_{\rm CMB}(z)/T_0 \right)}{\ln (1+z)}  \mathcomma
\label{eq:beta}
\end{equation}
where $\beta(z)$ is expressed as a function of the CMB temperature at a given redshift, $T_{\rm CMB}(z)$, and the present-day CMB temperature, $T_0$. Note that if the standard scaling given by \cref{eq:tz_scal} holds, $\beta$ is predicted to vanish identically at all redshifts: $\beta(z) = 0\ \forall z$. However, if this relation is violated at some point during the cosmic thermal history due to physical mechanisms that alter the adiabatic cooling of CMB photons, $\beta$ can capture such deviations by departing from the null value and possibly acquiring a redshift dependence, effectively becoming a function of redshift: $\beta = \beta(z)$. In this subsection, we perform a GP reconstruction to extract the shape of $\beta(z)$ across the redshift range covered by the data. This allows us to test whether $\beta(z)$ remains consistently compatible with zero and to constrain possible deviations from the standard scaling throughout the full redshift interval.

We summarise our results in \cref{fig:cc_data_recbeta}, where we present the reconstruction of the $\beta(z)$ function across the redshift range covered by the data. The individual $\beta$ data points (shown in blue for the SZ data and in green for the spectroscopic data) are obtained by translating the observed measurements $T_{\rm CMB}(z)$ listed in \cref{tab:combined_data} into measurements of $\beta(z)$ by means of \cref{eq:beta}. The associated uncertainties are consistently propagated from the original temperature errors. Using these $\beta(z)$ data points as input, we employ GP techniques to reconstruct $\beta(z)$. The reconstructed function is shown with confidence bands of 1$\sigma$, 2$\sigma$, and 3$\sigma$, depicted in progressively lighter shades of pink, as in previous figures. The overlaid data points illustrate the consistency between the observed $\beta(z)$ estimates and the GP-reconstructed curve.

The left panel of \cref{fig:cc_data_recbeta} focuses on the low-redshift regime ($0 < z < 1$), where the high precision of the SZ-based $T_{\rm CMB}(z)$ measurements allows for tight constraints on $\beta(z)$. In this range, most data points lie comfortably within the $1\sigma$ uncertainty band of the reconstruction and are individually compatible with it. The reconstruction itself stays broadly consistent with $\beta(z) = 0$, supporting the expected adiabatic cooling of the CMB and confirming the robustness of the standard scaling in the nearby Universe.

In the middle panel, we show the $\beta(z)$ reconstruction obtained using only the green spectroscopic measurements. As expected, the reconstruction is well constrained only in the redshift interval densely populated by data, roughly between $z \sim 1$ and $z \sim 3.5$. Outside this range, the lack of spectroscopic points both at low redshift and beyond $z \gtrsim 3.5$ leads to a significant broadening of the uncertainty bands. Still, the inclusion of the highest-redshift point around $z \sim 6$ ensures that the increase in uncertainty at high $z$ remains under control, and the overall reconstruction remains in excellent agreement with $\beta(z) = 0$, even when extrapolated to redshift intervals not directly sampled by the data.

The right panel shows the reconstruction obtained from the full dataset, combining SZ and spectroscopic measurements. In this case, the constraining power of the low-redshift SZ points improves the fit in the nearby Universe, tightening the bands at $z \lesssim 1$ compared to the spectroscopic-only case. It is also worth noting that the inclusion of a larger number of data points leads to a slightly more fluctuating reconstruction, which appears somewhat more sensitive to the noise present in the datasets. Nonetheless, this behaviour is not a concern as the individual $\beta(z)$ values remain in good agreement with the reconstructed curve, and the GP-based reconstruction continues to be fully consistent with $\beta(z) = 0$ across the entire redshift range.

Overall, we find no significant indication of deviations from the standard redshift scaling, and the reconstructed $\beta(z)$ remains consistent with the null value throughout the full interval probed by current data.

\subsection{$\chi^2$ Minimisation Analysis} \label{sec:chi2}

\begin{figure*}[htb]
    \centering
    \includegraphics[width=1\textwidth]{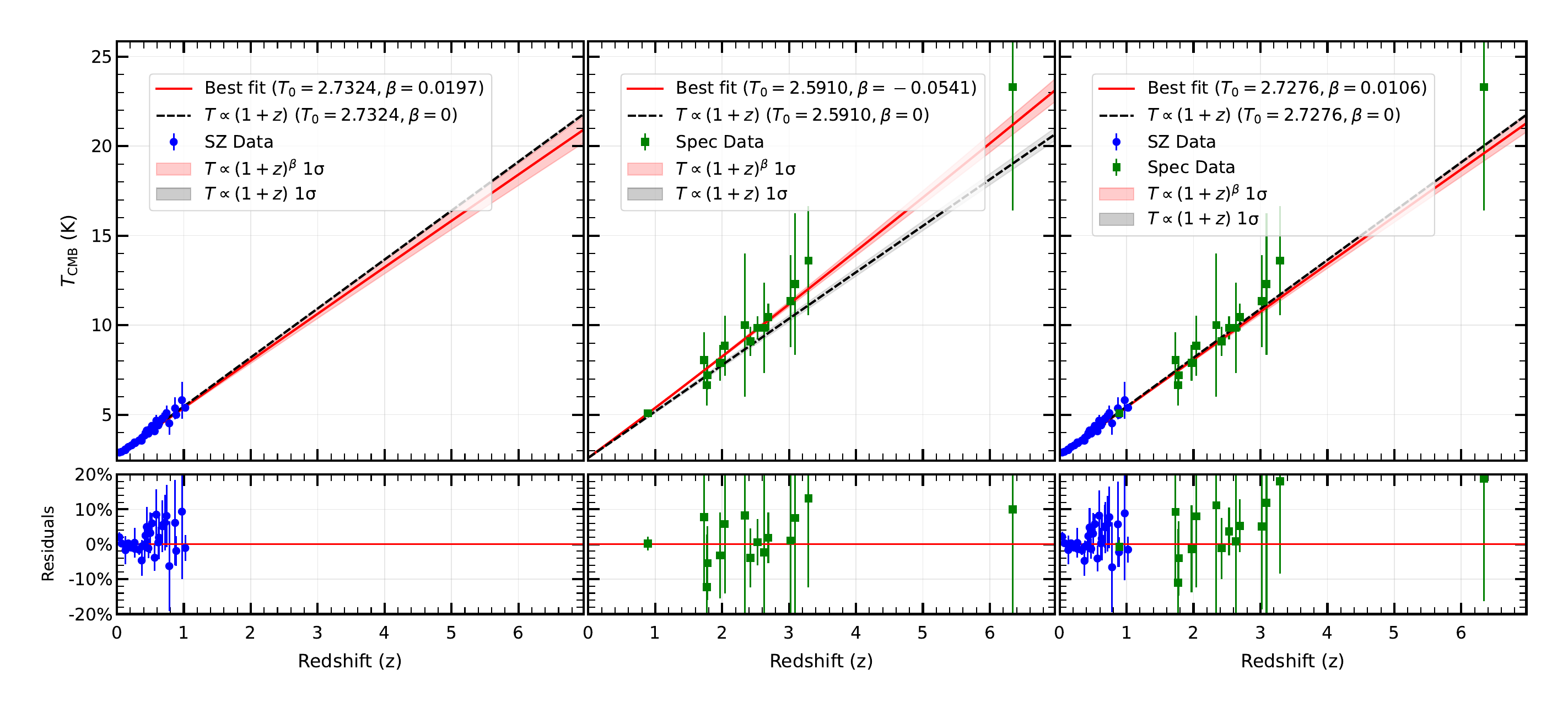}
    \caption{Comparison between the best-fit prediction for $T_{\rm CMB}(z)$ under the standard temperature-redshift evolution in~\cref{eq:tz_scal} (black) and the $\beta$-scaling power-law model in~\cref{eq:tzbeta_scal} (pink). The analysis is performed considering only the SZ data (left panel, blue circles), the spectroscopic measurements (middle panel, green squares), as well as their combination (right panel). The lower panels show the normalised residuals in each case.}
    \label{fig:cc_datachi}
\end{figure*}

The last step of our analysis is to investigate deviations from the standard scaling and derive updated constraints on $T_0$ and the deviation parameter $\beta$ through a $\chi^2$-minimisation procedure. As usual, we adopt \cref{eq:tzbeta_scal} to parametrise possible deviations from the standard scaling. However, the $\chi^2$-minimisation analysis performed in this subsection differs significantly from the GP reconstruction of $\beta(z)$ carried out in the previous subsection, both in terms of methods and aims. To prevent any misunderstanding, we stress once more that the GP reconstruction of $\beta(z)$ was intended to test the consistency of $\beta(z) = 0$ across the full redshift range, allowing for a redshift-dependent reconstruction of $\beta(z)$ directly from the data. Conversely, in this subsection, we assume $\beta$ to be a \textit{constant} parameter and derive the best-fitting values of $\beta$ and $T_0$ from a global fit to the different measurements of $T_{\rm CMB}(z)$ from SZ and Spec data, still adopting the parametrisation of \cref{eq:tzbeta_scal}. On the one hand, this approach allows us to constrain the parameter space and evaluate the statistical significance of the results more straightforwardly. On the other hand, by employing and comparing both methods, we can cross-validate our findings, providing a more comprehensive understanding of the data and establishing more robust constraints on potential deviations from the standard temperature-redshift relation across cosmic history.

In more quantitative terms, to determine the best-fit value of the model parameters $T_0$ and $\beta$, we minimise
\begin{equation}
\chi^2 (T_0,\beta) = \sum_{i} \frac{\left[T_{\text{obs}}(z_i) - T_{\text{th}}(z_i, T_0,\beta) \right]^2}{\sigma_{T_{\text{obs}}}^2(z_i)} \mathcomma
\label{eq:chi2}
\end{equation}
where $T_{\text{th}}(z_i, T_0, \beta)$ is the theoretical value expected for the CMB temperature at redshift $z_i$ for a given pair of values $T_0$ and $\beta$, as predicted by \cref{eq:tzbeta_scal}, while $T_{\text{obs}}(z_i)$ are the corresponding observed values -- with associated uncertainties $\sigma_{T_{\text{obs}}}(z_i)$ -- listed in \cref{tab:combined_data}. The index $i$ in the sum runs over all the data points employed.

Our results are summarised in \cref{fig:cc_datachi}. The top panels compare the CMB temperature evolution models across different data combinations. The red solid lines correspond to the best-fit model based on the functional form given in \cref{eq:tzbeta_scal}, with the associated $1\sigma$ confidence bands shown in shaded red. For comparison, the black dashed lines indicate the standard scaling prediction ($\beta = 0$), accompanied by its own $1\sigma$ uncertainty region in shaded black. The bottom panels display the normalised residuals, defined as the difference between the observed CMB temperature and the value predicted by \cref{eq:tzbeta_scal} at the same redshift (using the best-fit values for $T_0$ and $\beta$), normalised by the observational uncertainties. The horizontal red line at zero marks perfect agreement between data and model, with an unbiased model expected to yield residuals randomly scattered around this line.

In our $\chi^2$-minimisation analysis, we distinguish three different scenarios. First, we focus exclusively on low-redshift SZ data, with the results summarised in the left panel of \cref{fig:cc_datachi}. In this case, the best-fit parameters are $T_0 = 2.7324 \pm 0.0098\, \text{K}$  and $\beta = 0.0197 \pm 0.0155$. Notably, $\beta$ is consistent with $\beta = 0$ within two standard deviations, as indicated by the standard scaling (black line).

Next, we analyse spectroscopic data, as shown in the middle panel of \cref{fig:cc_datachi}. The best-fit parameters in this case are $T_0 = 2.5910 \pm 0.0443 \, \text{K}$ and $\beta = -0.0541 \pm 0.0208$. Taken at face value, the spectroscopic data suggest a best-fitting scale that deviates from the standard scaling, as we observe a significant difference between the best-fit values of $T_0$ when $\beta$ is allowed to vary freely and when it is fixed to $\beta = 0$ (\textit{i.e.}, comparing the black and red lines). However, as discussed in the previous section, focusing solely on spectroscopic data introduces limitations, particularly in terms of precision at low redshifts and in the range $z \in [3\,,\,6]$, where the number of data points decreases. Therefore, while we present this result for completeness, we emphasise that combining both SZ and spectroscopic data provides a more reliable approach. 

The results from this combined analysis are shown in the right panel of \cref{fig:cc_datachi} and represent our baseline results with the following best-fit parameters at the 68\% confidence level:
\begin{align}
T_0 &= 2.7276 \pm 0.0095\, \text{K} \mathcomma \nonumber \\
\beta &= -0.0106 \pm 0.0124 \mathperiod
\end{align}
In this case, we find excellent agreement with $\beta = 0$, with a deviation of the best-fit value of less than $1\sigma$.\footnote{For the sake of completeness, we report that the $\chi^2$ analysis of the SZ data yields $\chi^2 = 20.42$ with 28 degrees of freedom (dof) -- \textit{i.e.}, 30 SZ measurements minus 2 free parameters in the fit: $T_0$ and $\beta$. This corresponds to a normalised $\chi^2/\mathrm{dof} = 0.73$. When combining the SZ and spectroscopic datasets (45 data points in total), the best-fit yields $\chi^2 = 23.47$ with 43 dof, resulting in a reduced $\chi^2/\mathrm{dof} = 0.55$. The addition of 15 data points from the spectroscopic dataset increases the total $\chi^2$ by less than 1 per dof, thereby reducing the overall $\chi^2/\mathrm{dof}$. This is due to the large uncertainties in the spectroscopic measurements, which make them less influential in the fit than would be expected from their number of data points, reflecting their lower weight in the overall fit.}
These results further support the validity of the standard scaling relation ($\beta = 0$), limiting potential deviations due to non-standard processes that might cause non-adiabatic cooling or heating of CMB photons across cosmic evolution. Furthermore, the value we infer for $T_0$ is in remarkable agreement with direct measurements from FIRAS and ARCADE, well within one standard deviation.

\section{Discussions and Implications} \label{sec:discussion}

\subsection{Constraints on $T_0$ and physical implications} \label{sec:t0_const}

\begin{figure*}[htb]
    \centering
    \includegraphics[width=\textwidth]{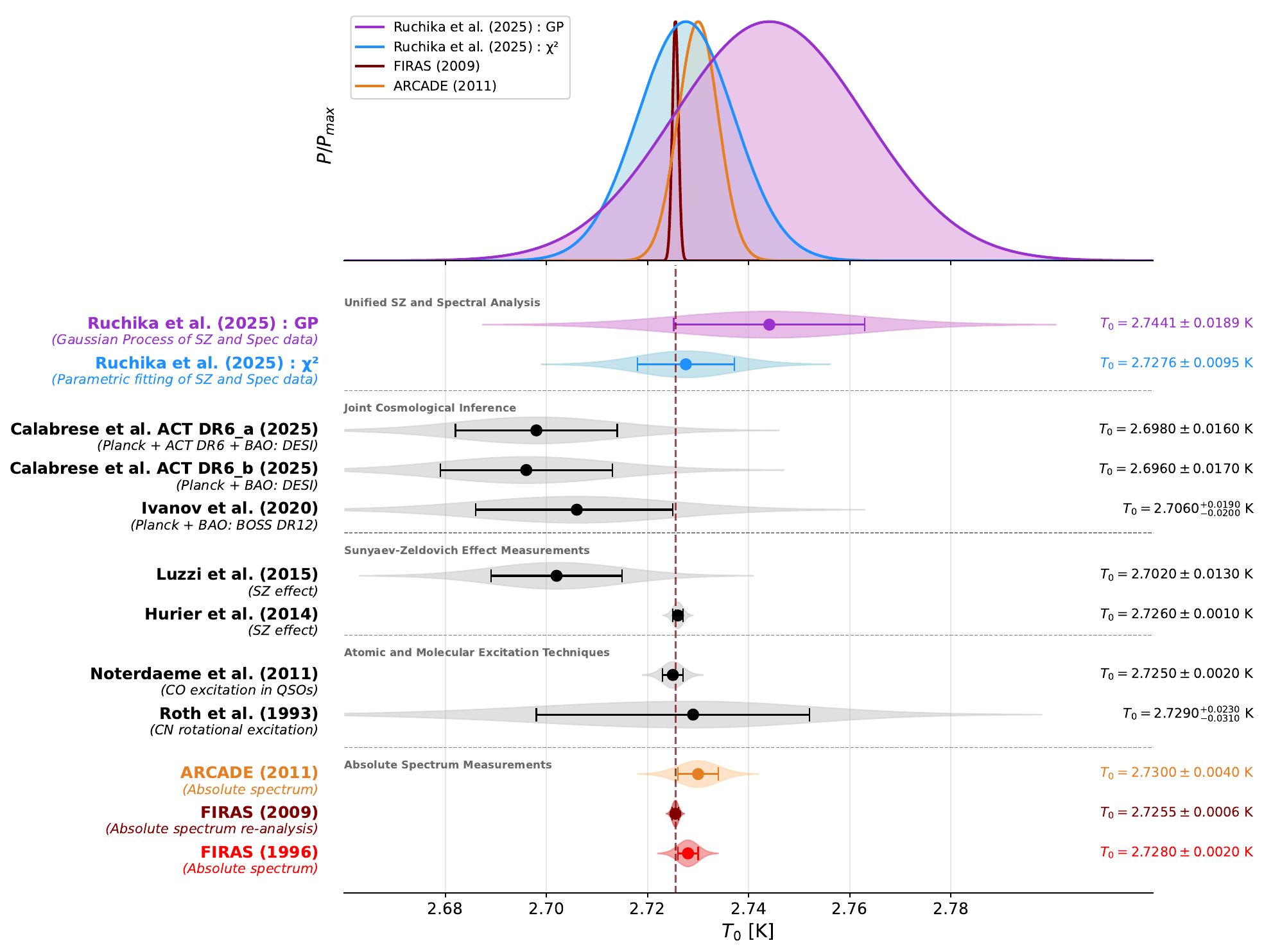}
    \caption{Whisker plot displaying several notable $T_0$ determinations from the literature, each obtained using different methodologies and datasets as described in the text. The results derived in this work are presented at the top of the figure under the label \textit{Ruchika et al. (2025)}, with the value inferred from GP reconstruction shown in purple and the one from the $\chi^2$ analysis shown in blue. The FIRAS and ARCADE results are represented in red and orange, respectively, at the bottom of the figure. For context, additional estimates from previous studies are shown in black. All uncertainties are quoted at the 68\% confidence level. The upper panel of the plot displays the full 1D posterior distributions of our results and the direct estimates from FIRAS and ARCADE for comparison.}
    \label{fig:T0_measurements}
\end{figure*}

This analysis provides two independent estimates of the present-day CMB monopole temperature, $T_0$, using complementary approaches. From the model-independent GP reconstruction of $T_{\rm CMB}(z)$, we obtain $T_0 = 2.744 \pm 0.019$ K. In parallel, from a $\chi^2$ minimisation analysis within the parametric framework defined by \cref{eq:tzbeta_scal}, we infer a best-fit value of $T_0 = 2.7276 \pm 0.0095$ K. Both estimates are broadly consistent with the leading direct measurements by FIRAS and ARCADE. This agreement is also clearly illustrated in \cref{fig:T0_measurements}.

\cref{fig:T0_measurements} shows a whisker plot collecting several notable $T_0$ determinations from the literature, each obtained through different methodologies and data. Our results are reported at the top of the figure under the label \textit{Ruchika et al. (2025)}. We show in purple the value inferred by GP reconstruction and in cyan the one coming from the $\chi^2$ analysis. These can be directly compared to the FIRAS and ARCADE results, shown at the bottom in red and orange, respectively.\footnote{For FIRAS, we include both the original 1996 result and the updated 2006 analysis of full spectrum.} For these cases, we also display the full 1D posterior distributions in the upper panel of the plot.

For context, in \cref{fig:T0_measurements}, we also include additional estimates from previous studies. In the ``Atomic and Molecular Excitation Techniques'' block, we report two values obtained exclusively through these approaches: a much earlier determination from 1993 \cite{Roth:1993ApJ} and a more recent one from 2011 \cite{Noterdaeme:2010tm}. In the ``Sunyaev-Zeldovich Effect Measurements'' block, we show $T_0$ estimates from 2014 \cite{Hurier:2013ona} and 2015 \cite{Luzzi:2015via} derived using SZ data alone, each relying on different assumptions and fitting strategies. Our analysis, in contrast, combines updated data from both methodologies (atomic and molecular excitation lines at high redshift and SZ measurements at low redshift), allowing us to bridge the two regimes in a coherent framework. This comparison highlights not only how the estimates have evolved with the latest data, but also the value of combining SZ and spectroscopic information. Overall, we find good consistency with previous results derived from similar techniques. Notably, our estimates are not only in excellent agreement with direct measurements but also fall well within the range spanned by earlier literature. One key difference, however, lies in the model-independence of our GP-based inference. Since it does not rely on any parametric form for $T(z)$, the resulting constraint on $T_0$ holds true in any model that could alter the value of $T_0$ inferred either directly or indirectly from data, whether under standard scaling or modifications thereof.

Finally, for completeness, we also report more recent determinations from ``Joint Cosmological Inference'', in which $T_0$ is promoted to an effective seventh parameter within $\Lambda$CDM and inferred alongside other cosmological parameters. These estimates are again broadly consistent with our findings, though one might speculate about a slight tendency toward lower $T_0$ values (\textit{i.e.}, leftward shift on the $x$-axis in the figure). As discussed in the introduction and first shown in Ref.~\cite{Ivanov:2020mfr}, this shift can become significantly more pronounced when applying a prior on the Hubble constant from SH$_0$ES together with the \textit{Planck} temperature and polarisation anisotropy data, leading to $T_0 = 2.5640^{+0.0510}_{-0.0490}$ K, in $\sim 3\sigma$ tension with direct measurements of $T_0$. This is due to the unavoidable degeneracy between $T_0$ and $H_0$: from the CMB-only perspective, $T_0$ can effectively quantify the total expansion from recombination to today. Lower values of $T_0$ recast into changes in the angular diameter distance to the last scattering surface, an effect that mimics the impact of changing $H_0$ almost perfectly. In this context, both our $T_0$ estimate from $\chi^2$ minimisation and the model-free GP reconstruction disfavour this shift toward smaller $T_0$ values, instead showing solid agreement with the absolute spectral measurements from FIRAS and ARCADE. In this sense, and bearing in mind the fully model-independent GP reconstruction of $T_{\rm CMB}(z)$, our results also provide stringent constraints on any non-standard cosmological model that attempts to increase $H_0$ via modifications to the $T_{\rm CMB}(z)$ relation.

\subsection{Constraints on $\beta$ and physical implications} \label{sec:beta_const}

In the second part of our analysis, we adopted the parametric form given in \cref{eq:tzbeta_scal}, which introduces an additional degree of freedom, $\beta$, expected to vanish identically in the case of standard adiabatic evolution. While this parameter allows for possible departures from adiabaticity in the thermal history of CMB photons, both our GP reconstruction and $\chi^2$ minimisation yield results consistent with the standard scaling. The former approach promotes $\beta(z)$ to an effective redshift-dependent function, showing that it remains consistent with zero across the full range covered by the data. The latter treats $\beta$ as a constant free parameter, yielding a constraint of $\beta = -0.0106 \pm 0.0124$. These findings effectively constrain a broad class of physical and phenomenological mechanisms that would otherwise induce deviations from the standard relation. In what follows, we provide examples of such scenarios, illustrating how both sets of results can place meaningful limits on a variety of well-motivated physical models, many of which are of interest in the context of cosmological tensions and fundamental physics.
\begin{itemize}
    \item Restricting our attention to the simplest scenario in which the temperature–redshift relation follows the parametrisation of \cref{eq:tzbeta_scal} with a constant $\beta$, we note that this is already sufficient to capture the leading effects of a broad class of mechanisms that can alter the adiabatic evolution of the CMB or modify the geodesic propagation of CMB photons. Interestingly, within specific theoretical frameworks, such deviations from the standard scaling can be directly connected to violations of the DDR, which links the luminosity distance $d_L$ and the angular diameter distance $d_A$ \textit{via} $d_L(z) = d_A(z)(1+z)^2$ and relies on photon number conservation and the validity of metric theories of gravity.\footnote{For a broad range of works over the past two decades discussing the possibility of testing the DDR against a variety of astrophysical and cosmological datasets, both in model-dependent and model-independent frameworks, the interested reader can refer to Refs.~\cite{Uzan:2004my,DeBernardis:2006ii,Holanda:2010ay,Holanda:2010vb,Li:2011exa,Nair:2011dp,Liang:2011gm,Meng:2011nt,Holanda:2011hh,Khedekar:2011gf,Goncalves:2011ha,Lima:2011ye,Holanda:2012at,Cardone:2012vd,Yang:2013coa,Zhang:2014eux,Santos-da-Costa:2015kmv,Wu:2015ixa,Wu:2015ixa,Liao:2015uzb,Rana:2015feb,Ma:2016bjt,Holanda:2016msr,More:2016fca,Rana:2017sfr,Li:2017zrx,Lin:2018qal,Qi:2019spg,Holanda:2019vmh,Zhou:2020moc,Qin:2021jqy,Bora:2021cjl,Mukherjee:2021kcu,Liu:2021fka,Renzi:2021xii,Tonghua:2023hdz,Qi:2024acx,Yang:2024icv,Tang:2024zkc,Jesus:2024nrl,Favale:2024sdq,Qi:2024acx,Alfano:2025gie,Yang:2025qdg,Keil:2025ysb,Xu:2022zlm} while for projected constraints expected from future surveys, see, \textit{e.g.}, Refs.~\cite{Cardone:2012vd,Yang:2017bkv,Fu:2019oll,Hogg:2020ktc,Renzi:2020bvl,Euclid:2020ojp}.} Departures from this relation are typically parametrised as $d_L(z) = d_A(z)(1+z)^{2+\epsilon}$~\cite{Euclid:2020ojp}, where $\epsilon$ quantifies the deviation from the canonical scaling.  Although, in general, $\epsilon$ and $\beta$ describe distinct classes of new physics, under specific assumptions, deviations from the DDR can induce corresponding changes in the redshift evolution of the CMB temperature, connecting the two parameters. For instance, assuming for simplicity that these deviations are achromatic (\textit{i.e.}, independent of photon wavelength) and approximately adiabatic (so that the CMB spectrum remains close to a blackbody), a power-law breaking in the DDR can propagate into modifications of the CMB temperature evolution, parametrised as in \cref{eq:tzbeta_scal}, thereby establishing the direct connection $\epsilon = -\frac{3}{2} \beta$, see, \textit{e.g.}, Ref.~\cite{Avgoustidis:2011aa}. Within these specific theoretical frameworks, our constraints on $\beta$ can be translated into bounds on a subset of DDR-violating scenarios, providing an independent and complementary test of this fundamental relation and constraining
    \begin{equation}
    d_L(z)/d_A(z) \sim (1+z)^{2.0159\pm0.0186} \mathperiod
    \label{eq:DDR_limits}
    \end{equation}
    Interestingly, in Ref.~\cite{Teixeira:2025czm}, some of us showed that a phenomenological model featuring a broken power-law DDR, $d_L(z)/d_A(z) \sim (1+z)^{1.866}$ at low redshifts, combined with a constant phantom equation of state $w \sim -1.155$, can explain the Hubble tension. The aforementioned analysis was a proof of concept based solely on geometric observables, such as Type Ia supernovae and BAO, and \textit{did not} require any modifications to the CMB temperature–redshift relation, thereby remaining unaffected by the limits discussed above. However, should such phenomenological deviations be implemented within a physical mechanism, care must be taken to ensure that these models do not also affect photon propagation in a way that modifies $T_{\rm CMB}(z)$, as described above. Otherwise, they would fall within the class of models constrained by \cref{eq:DDR_limits} from our present analysis, which would then severely restrict any deviation from the DDR. This highlights a key consistency requirement: models that alter the DDR \textit{via} a power-law breaking must either decouple from the thermal evolution of the CMB or be constructed in a way that remains compatible with precise $T_{\rm CMB}(z)$ measurements, underscoring the complementarity between geometric and thermodynamic probes in testing departures from standard cosmology.
  
    \item As discussed in \cref{sec:tz_dev}, promoting the parameter $\beta$ to a redshift-dependent function $\beta(z)$ and reconstructing its evolution using GP techniques shows that $\beta(z)$ remains consistently compatible with $\beta(z) = 0$ across the entire redshift range probed by data. This result supports the standard adiabatic cooling of the CMB and places stringent bounds on a broad class of scenarios that predict a non-standard redshift evolution of the CMB temperature. A particularly illustrative example is provided by models in which photon number is not conserved. In these scenarios, the comoving photon number density evolves with redshift as $N(z)$, rather than remaining constant at $N_0$ as in standard cosmology. This modifies the temperature–redshift relation of the CMB cooling to~\cite{Avgoustidis:2011aa}
    \begin{equation}
    T_{\rm CMB}(z) = T_0(1+z)\left(\frac{N(z)}{N_0}\right)^{1/3} \mathperiod
    \label{eq:tzN_scal}
    \end{equation}
    This scaling, and by extension, these models, can be naturally captured by \cref{eq:tzbeta_scal} once $\beta(z)$ is allowed to vary with redshift. Specifically, the deviation function $\beta(z)$ in \cref{eq:tzbeta_scal} can be expressed in terms of the photon number evolution $N(z)$ from \cref{eq:tzN_scal} as
    \begin{equation}
    \beta(z) = -\frac{1}{3} \left[ \frac{\ln\left( N(z)/N_0\right) }{\ln(1+z)} \right] \mathcomma
    \end{equation}
    where $\beta(z) = 0$ corresponds to standard adiabatic evolution with conserved photon number ($N(z) = N_0$), and deviations from $\beta (z) = 0$ indicate photon non-conservation. Positive values of $\beta(z) > 0$ suggest a decrease in photon number with redshift ($N(z) < N_0$), potentially due to photon absorption or conversion into other particles. In contrast, negative values of $\beta(z) < 0$ imply an increase in photon number ($N(z) > N_0$), possibly as a result of exotic particle decays or photon production mechanisms such as axion-photon mixing. Within this framework, our reconstructed constraints on $\beta(z)$, shown in \cref{fig:cc_data_recbeta}, can be directly translated into limits on deviations from $N(z)/N_0=1$, providing a model-independent probe of mechanisms that alter photon number across cosmic history. The observed consistency with $\beta(z) \approx 0$ at all redshifts supports photon number conservation.

    \item Other physically meaningful implications of our constraints on $\beta$ (whether treated as a constant or as a redshift-dependent quantity) concern models featuring a varying speed of light. In such scenarios, the thermal history of the CMB deviates from the standard scaling $T_{\rm CMB}(a) \propto c^2/a$, where $a = 1/(1+z)$ is the scale factor~\cite{Barrow:1998eh}. For instance, if $c$ evolves with time as $c(t) \propto (\ln t)^{-1}$ or follows a power-law behaviour, the temperature–redshift relation is correspondingly altered. In these frameworks, the modified scaling described by \cref{eq:tzbeta_scal} can effectively encapsulate such deviations, with $\beta$ mapping onto specific functional forms of $c(z)$. Depending on the underlying theory, smoothly varying $c(z)$ may correspond to a non-trivial evolution of $\beta(z)$, while other scenarios may be well described by a constant $\beta$. This connection ties our empirical results to more fundamental physics, as variations in $c$ would also impact the fine-structure constant $\alpha = e^2/\hbar c$~\cite{Uzan:2010pm,Tohfa:2023zip}, potentially affecting early-universe dynamics through modifications to gauge couplings and other interactions governed by dimensionless constants.
    
\end{itemize}

\section{Conclusions}\label{sec:conc}

In this work, we studied the evolution of the CMB temperature, $T_{\rm CMB}(z)$, and investigated potential deviations from the standard adiabatic scaling relation, \cref{eq:tz_scal}. We explored the latest Sunyaev-Zeldovich (SZ) effect measurements and molecular line excitation (spectroscopic) data -- which combined span a redshift range of $0 \lesssim z \lesssim 6$ -- following different independent (yet complementary) approaches:

\begin{itemize} 

\item We first reconstructed the redshift evolution of $T_{\rm CMB}(z)$ in a model-independent way using Gaussian Process (GP) regression. Our findings, summarised in \cref{fig:cc_data32}, show that the low-redshift SZ-based reconstruction remains within two standard deviations from the standard scaling across the explored range, demonstrating good consistency. At higher redshifts, reconstructions based on spectroscopic data exhibit broader uncertainty bands, reflecting the larger measurement errors. Overall, $T_{\rm CMB}(z)$ remains in good agreement with the standard adiabatic scaling across the full redshift range when both datasets are combined. In this case, we infer a present-day CMB monopole temperature of $T_0 = 2.744 \pm 0.019$ K, which is in good agreement with the direct measurements from FIRAS and ARCADE within one standard deviation.

\item We then further investigated potential deviations from the standard scaling by introducing a modified temperature-redshift relation, described by \cref{eq:tzbeta_scal}, which brings in an additional degree of freedom $\beta$ (predicted to be identically null in the standard case) that can capture possible departures from adiabaticity during the thermal evolution of CMB photons. We first performed a GP reconstruction of $\beta(z)$ directly from the data to test its potential evolution across the redshift range probed by observations and assess its consistency with the null case. As shown in \cref{fig:cc_data_recbeta}, our analysis demonstrates that $\beta(z)$ remains consistent with zero within the $2\sigma$ confidence interval throughout the full redshift range covered by data, supporting the validity of the standard adiabatic cooling assumption. We found that low-redshift ($z < 1$) SZ measurements offer tighter constraints, while larger uncertainties appear at higher redshifts ($z > 3$) from spectroscopic data.

\item Finally, we performed a $\chi^2$-minimisation analysis to derive updated constraints on $T_0$ and the deviation parameter $\beta$ in \cref{eq:tzbeta_scal}. Here, we assumed $\beta$ to be constant. Combining SZ and spectroscopic data, we obtained best-fit values of $T_0 = 2.7276 \pm 0.0095$ K and $\beta = -0.0106 \pm 0.0124$. Importantly, $\beta$ remains compatible with zero, indicating once again no statistically significant deviation from the standard adiabatic evolution of the CMB temperature. Furthermore, $T_0$ remains always in perfect agreement with the direct measurements from FIRAS and ARCADE at the 1$\sigma$ level.
\end{itemize}

Overall, our results are consistent with previous studies but provide a more comprehensive assessment by examining the possible redshift dependence of the deviation parameter $\beta$ and by testing the CMB cooling in a model-independent manner employing GP techniques.

A potential caveat surrounding our conclusions lies in the fact that the parametrisation in \cref{eq:tzbeta_scal}, while adequate at low redshifts ($z \lesssim 1$), may become unreliable at higher redshifts. This limitation arises because physically motivated models typically predict different temperature evolution behaviours during the matter-dominated era. Our choice to adopt this parametrisation stems primarily from its widespread use in the literature. However, future investigations should explore more physically motivated models, particularly as observational data continue to improve in quality, coverage, and redshift reach.

Very significant improvements are expected in the coming years. From the theoretical side, there is significant interest in studies investigating the link between discrepancies in the CMB and additional signatures that could help explain them. One interesting example involves detailed modelling of non-standard recombination physics, exploring potential energy injections or modifications to the baseline models~\cite{Lynch:2024gmp,Lynch:2024hzh,Jedamzik:2023rfd,Mirpoorian:2024fka}. These deviations have been suggested as possible solutions to reduce the Hubble tension while potentially affecting the thermodynamics of the CMB. In parallel, growing attention has been given to anisotropies in CMB spectral distortions -- deviations from a perfect blackbody spectrum that can arise from non-standard energy release in the early Universe, see, \textit{e.g.}, Ref.~\cite{Chluba:2025wxp} for a recent overview. These distortions, including $\mu$- and $y$-type signatures, can provide additional insights into the thermal history of the Universe and are sensitive to both early-time physics and late-time processes, such as structure formation. Their potential angular dependence has been suggested as a complementary probe to the usual CMB temperature and polarisation anisotropies, potentially shedding light on unresolved tensions and revealing hints of new physics~\cite{Chluba:2022xsd,Chluba:2022efq,Kite:2022eye}. Another avenue has examined the role of peculiar velocity fields and their imprint on the CMB anisotropy spectra, particularly in relation to large-scale anomalies such as the dipole asymmetry \cite{Dalang:2021ruy,Secrest:2020has,Secrest:2022uvx,Abghari:2024eja}. From the observational side, the next generation of space-based CMB missions, including proposed projects like CMB-HD \cite{CMB-HD:2022bsz} and the Probe of Inflation and Cosmic Origins (PICO) \cite{NASAPICO:2019thw}, could substantially increase the number of available SZ measurements of the CMB temperature. For spectroscopic measurements, ALMA \cite{Walter_2016} has already begun making valuable contributions, while ESPRESSO \cite{2021A&A...646A.158T} on the VLT \cite{2024GCN.37369....1S} is now fully operational and producing high-quality data. The upcoming ANDES (formerly HIRES) instrument planned for the European Extremely Large Telescope (ELT) \cite{2024SPIE13096E..13M} promises even greater precision in the next decade. Recent advances in CO, CI, and molecular line observations at high redshifts have improved our ability to probe the thermal history of the Universe. Future work combining these next-generation facilities with improved high-redshift observations will further refine constraints on $T_{\rm CMB}(z)$, potentially uncovering new physics beyond the standard cosmological model.

\section*{Acknowledgements}
R is supported by Project SA097P24, funded by Junta de Castilla y Leon. The author R was also supported by "Theoretical Astroparticle Physics" (TAsP), iniziativa specifica INFN.
WG\ acknowledges support from the Lancaster–Sheffield Consortium for Fundamental Physics through the Science and Technology Facilities Council (STFC) grant ST/X000621/1.
EMT is supported by funding from the European Research Council (ERC) under the European Union's HORIZON-ERC-2022 (grant agreement no. 101076865).
The work of AM was partially supported by the research grant number 2022E2J4RK "PANTHEON: Perspectives in Astroparticle and Neutrino THEory with Old and New messengers" under the program PRIN 2022 funded by the Italian Ministero dell'Universit'a e della Ricerca (MUR) and by the European Union – Next Generation EU, as well as by the Theoretical Astroparticle Physics (TAsP) initiative of the Istituto Nazionale di Fisica Nucleare (INFN).

\bibstyle{unsrt}
\bibliography{main} 

\end{document}